\documentstyle[twocolumn,aps,prl,psfig19]{revtex}
\def\s{\sqrt{S_2(r)}}
\def\av#1{{\left\langle#1\right\rangle}}
\begin{document}
\wideabs{
\draft
\preprint{ }
\title{%
Probability Density Function of Longitudinal Velocity Increment\\
in Homogeneous Turbulence}
\author{%
Naoya {\sc Takahashi},$^{1}$
Tsutomu {\sc Kambe},$^{1}$ Tohru {\sc Nakano},$^{2}$ \\
Toshiyuki {\sc Gotoh},$^{3}$ and Kiyoshi {\sc Yamamoto}$^{4}$}
\address{%
$^1$ Department of Physics, University of Tokyo, Tokyo 113-0033,\\
$^2$ Department of Physics, Chuo University, Tokyo 112-8551,\\
$^3$ Department of System Engineering,
Nagoya Institute of Technology, Nagoya 466-8555,\\
$^4$ National Aerospace Laboratory, Chofu, Tokyo 182-8522}
\date{\today}
\maketitle

\begin{abstract}
Two conditional averages for the longitudinal velocity increment $u_r$ in
the simulated turbulence are estimated: $h(u_r)$ is the average of
the difference of the Laplacian of the velocity field
 with the $u_r$ value fixed,
while $g(u_r)$ is the corresponding one
of the square of the difference of the velocity gradient.
Based on the physical argument, we suggest the fitting formulae for
$h$ and $g$,
which quite satisfactorily reproduced to the $512^3$ DNS data.
The predicted PDF is characterized as
    (1) the Gaussian distribution for smaller amplitudes,
    (2) the exponential distribution for larger ones,
and (3) a prefactor before the exponential function for the intermediate ones.
\end{abstract}
\pacs{PACS numbers: 47.27.Gs}
\sloppy
}
\narrowtext
Probability density function (PDF) is one of the most
interesting subjects in turbulence\cite{rf:my}.
The  intermittent effect often referred to 
is reflected in the deviation of the PDF from the
Gaussianity.  In particular  the PDF for longitudinal velocity increments
has been the main concern in turbulence\cite{rf:cg,rf:po,rf:cm,rf:za}.
For large separation $r$ the PDF is Gaussian and it becomes
deviated more as $r$ decreases.
The PDF of its derivatives is even flatter than
the exponential distribution\cite{rf:sz,rf:vm,rf:sd,rf:bm}.
In the present letter we propose
a simple form of PDF, which is consistent with the observation so far.

The conditional average method\cite{rf:sy,rf:c,rf:a} is very useful to
understand the nature of turbulence.
In a previous paper\cite{rf:t} the method
is applied to the study of the passive scalar field advected by Navier-Stokes
turbulence. In the present letter the method is applied to the analysis of
statistics of the longitudinal velocity difference  at two points separated by $r$:
\(     U_r({\bf x}) \equiv U({\bf x}+r\hat{\bf x})- U({\bf x})\),
where $U({\bf x})$ is the $x$ component of the velocity field.
The first conditional average we study is
\begin{equation}
    H(U_r)=\av{\nabla_{{\bf x}}^2 U_r({\bf x})|U_r},
\end{equation}
where the average is taken with a value of $U_r$ fixed.  The second average is
\begin{equation}
    G(U_r)=\av{|\nabla_{{\bf x}} U_r({\bf x})|^2|U_r}.
\end{equation}

For later convenience $U$, $H$, and $G$ are made dimensionless as
\begin{eqnarray}
&&   u_r=\frac{U_r}{\s}, \quad h(u_r)=\frac{2 \nu \s}{J_2(r)}H(U_r),
                                                         \nonumber\\
&&  g(u_r)=-\frac{2 \nu}{J_2(r)} G(U_r),
\end{eqnarray}
where $\nu$ is the kinematic viscosity, $S_2(r)$ 
$=\av{U_r^2}$, 
and $J_{2}(r)=-2 \nu \av{|\nabla_{{\bf x}} U_r({\bf x})|^2}$.

A useful relation concerning the PDF $P(u)$ of $u$ is derived for a
statistically homogeneous system in terms of $h$ and $g$ as
follows\cite{rf:sy,rf:t,rf:ck}:
\begin{equation}
   P(u)g(u)=C \exp \left[ -\int^{u}_0 \frac{h(x)}{g(x)} dx \, \right]
    \label{P}
\end{equation}
where $C$ is a numerical constant.
This relation was numerically confirmed using our DNS data.

To begin with, let us derive possible forms of $h(u)$ and $g(u)$.
The equation for $U_r$ is formally written as
\begin{equation}
   \frac{\partial U_r}{\partial t}+T(U_r)=\nu \nabla^2 U_r, \label{t}
\end{equation}
where $T(U_r)$ is the inertial term containing the pressure effect.
Multiplying the above equation by $U_r$ we take the conditional average with
$U_r$ fixed:
\begin{equation}
   \frac{1}{2} \av{\frac{\partial U_r^2}{\partial t}\Bigm|U_r}
      +U_r \av{T(U_r)|U_r}
     =\nu U_r \av{\nabla^2 U_r|U_r}.
\end{equation}
In the quasi-steady state the first term on the left hand side can be
neglected.
Then
\begin{equation}
     \nu U_r H(U_r)=U_r \av{T(U_r)|U_r},
\end{equation}
which implies that $\nu U_r H(U_r)$ balances the energy transfer rate
in the inertial region.    It would take the form
\begin{equation}
   U_r \av{T(U_r)|U_r} \sim \frac{U_r^2}{\tau_r},  
\end{equation}
where the relaxation time $\tau_r$ would be given by the eddy turnover
time $r/U_r$;
then $U_r \av{T(U_r)|U_r} \sim U_r^3/r$.
Hence
\begin{equation}
  \nu H(U_r) \sim \frac{U_r^2}{r}\propto U_r^2.  \label{h1}
\end{equation}
In the inertial range, the eddy turnover time $\tau_r$ is given by the
Kolmogorov scaling form $\tau_r \sim r^{2/3}\overline{\varepsilon}^{-1/3}$
($\overline{\varepsilon}$ is the average dissipation rate),
independent of $U_r$, which implies that
\begin{equation}
    \nu H(U_r) \sim \varepsilon^{\frac{1}{3}}r^{-\frac{2}{3}}U_r \propto U_r .  \label{h2}
\end{equation}
The scaling (\ref{h1}) will be dominant for large values of
$U_r$ and the Kolmogorov scaling (\ref{h2}) is for small values of $U_r$.
Since  mechanisms coexist, a combined form
\begin{equation}
   h(u)=c_0 u+c_1 u^2  \label{h}
\end{equation}
is expected in the nondimensional form.
We have attempted a linear fitting for small values of $u$
and a quadratic one for large values of $u$,
but the fitting (\ref{h}) over the entire region is more satisfactory.
As the Reynolds number increases, the range proportional
to $u^2$ becomes wider.

Next we consider $G(U_r)$.  Multiplying (\ref{t}) by $U_r$
and taking a certain average, we have
\begin{equation}
    \av{U_r T(U_r)}=\nu\av{\nabla \cdot (U_r \nabla U_r)}
     -\nu\av{|\nabla U_r|^2}.
\end{equation}
The first term on the right hand side vanishes because of
the statistical homogeneity.
Then
\begin{displaymath}
  \int U_r \av{T(U_r)|U_r} P(U_r) dU_r
   =-\nu\int G(U_r) P(U_r) dU_r.
\end{displaymath}
Thus $G(U_r)$ is the conditional dissipation rate defined after the spatial
average being taken. Consider the extreme case where the spatial mixing is
strong.
Then the conditional dissipation rate $G(U_r)$ should be
independent of $U_r$.
Another limit is that $G(U_r)$ must be proportional
to the energy density $U_r^2$.
Hence one may expect a nondimensionalized form $g(u)=a_0+ a_2 u^2$.
(Since the spatial transfer works to diminish the $u^2$ scaling,
$a_2$ is expected to decrease as $r$ increases.)
Such a fitting is considerably good, but not excellent.
If one adds a linear term as
\begin{equation}
    g(u)=a_0 +a_1 u+a_2 u^2,  \label{g}
\end{equation}
the agreement is remarkable as we see later.
A linear term is necessary for
a smooth transition from the homogeneous to the intermittent dissipation.

Now we turn to $h(u)$ and $g(u)$ based on the DNS data and the
discussion of the PDF.
The simulations were carried out with resolutions of $256^3$
and $512^3$ mesh points using the NAL Numerical Wind Tunnel
for a couple of Reynolds numbers\cite{rf:yama}.
Since most of the results obtained are very similar to each other,
we present only the results of the run of $512^3$.
The turbulence decays from the initial spectrum $k^4 \exp(-2k^2)$ with no
forcing and the data  were processed when the energy dissipation rate
reaches the largest value.
 At that time the Taylor microscale Reynolds number is 120.
The inertial region in which $\av{U_r^3}/(\overline{\varepsilon}r)$
is nearly constant is located in $14 < r/\eta <51$, $\eta$ being
the Kolmogorov microscale.

The three functions, $h(u)$, $g(u)$, and $P(u)g(u)$ are estimated by
using the simulation data which are presented in the following.

\begin{figure}
\psfig{file=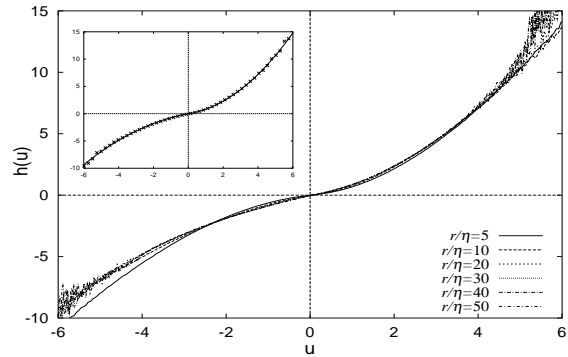,width=75mm,height=48mm}
\caption{$h$ vs $u$ for various values of $r$.
The solid lines in the inset represent the fitting for $r/\eta=20$,
the crosses being the data points.}
\label{fig:1}
\end{figure}

\noindent
(i) $h(u)$.
The curves of $h(u)$ against $u$ are depicted
for various values of $r$ in Fig.~\ref{fig:1}.
The interesting points are noticed by an inspection.
(a) All the curves for the inertial range separations fall on an
almost single line, which indicates that the conditional expectation of the
energy transfer rate is independent of the scale.
(b) $h(u)$ is not antisymmetric; in fact
the absolute magnitude is larger for large values of $u$ in the positive
region than in the negative region.
This tendency is obvious, because the longitudinal component
is similar to Burgers turbulence\cite{rf:y1},
where the positive slope region ($u>0$) works
on the negative slope to make it steeper,
which implies that the positive region transfers energy
to the negative slope side.
The excess of $h(u)$ in the positive region represents
a faster decay of the PDF there.
(c) $h(0)$ is slightly
negative for the inertial separations;
$h(0)=-0.05$ for $r/\eta=20$.
This is responsible for the skewed PDF.

Now we fit the formula (\ref{h}) with the data.
Since the curves are not symmetric,
we make a separate fit for $u>0$ and $u<0$.
The fitting is  remarkable except for the largest values of $|u|$,
where the data are scattered because of the insufficient number of data.
The inset in Fig.~\ref{fig:1} is an example of the curve for $r/\eta=20$
fitted in the interval $|u| \leq 8$,
where the solid line is the fitting function.
The fitting parameters are as follows:
For $r/\eta=5$ $c_0=0.67$, $c_1=0.27$ in the region $u>0$,
and $c_0=0.56$, $c_1=-0.20$ in the region $u<0$.
On the other hand, for $r/\eta=20$ in the inertial region,
$c_0=0.47$, $c_1=0.33$ for $u>0$, and $c_0=0.44$,
$c_1=-0.20$ for $u<0$.

\begin{figure}
\psfig{file=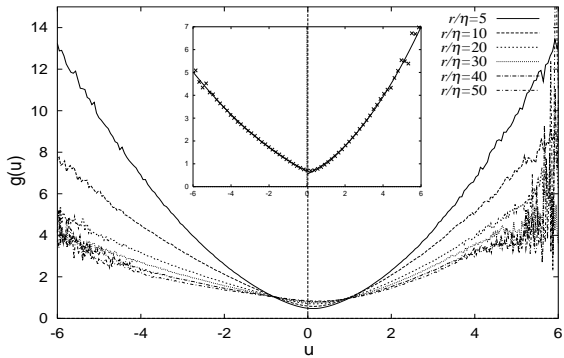,width=75mm,height=48mm}
\caption{$g$ vs $u$ for various values of $r$.
 The solid lines in the inset represent the fitting functions
 in the range (I) and (II) for  $r/\eta=20$,
 the crosses being the data points.
 The discrepancy for small amplitudes is seen.
 Those components are fitted by another quadratic expression.
}
\label{fig:2}
\end{figure}

\noindent
(ii) $g(u)$.
The curves of $g(u)$ against $u$ are given for various values
of $r$ in Fig.~\ref{fig:2}.
(a) The curves depend on $r$,
    which is responsible for the scale dependence of the PDF.
(b) As $r$ increases, $g(u)$ becomes flatter and $g(0)$ increases.
(c) $g(u)$ fluctuates considerably for large amplitudes;
    in particular, more conspicuous for large values of $r$.
    The reliable range is wider for $u<0$ than for $u>0$.

The formula (\ref{g}) is fitted to the data separately for the regions
(I) $u \geq b_1$, (II) $u \leq b_2$, and (III) $b_1> u> b_2$;
where $b_1=0.3$
and $b_2=-0.2$ for $r/\eta=20$, for instance.
(If only two regions $u>0$ and $u<0$ were chosen,
there would be a small discrepancy near $u = 0$.)
The fitting is quite satisfactory.
The solid line in the inset of Fig.~\ref{fig:2}
represents the fitting functions for $r/\eta=20$, where the fitting  is
processed for $-5 \leq u \leq -0.2$ and $5 \geq u \geq 0.3$.

In order to ascertain that the present result is consistent with the
observation\cite{rf:po,rf:cm},
which reports the scale dependence of the factor $c_1/a_2$,
we listed parameters $a_0$, $a_1$ and $a_2$ for
various values of $r$ in Table \ref{table:1}.
Since the region (II) is wider than the  region (I),
only parameters fitted for the region (II) are given;
$a_0$ is found to increase steadily with $r$.
The coefficient $a_2$ decreases with $r$ in the viscous and lower inertial ranges,
but the decrease cannot be seen for larger values.
Note that the determination of $a_2$ is a little subtle,
because the region exhibiting a quadratic dependence of $g(u)$
cannot be appreciated  with increasing $r$
in the simulations of low Reynolds numbers.
This scale dependence of $a_2$, however,
does not contradict with the observation\cite{rf:po,rf:cm}.
According to Ref.\cite{rf:po} $c_1/a_2 \sim r^{0.17}$,
meaning that $a_2 \sim r^{-0.17}$ because $c_1$ is independent of $r$
in the inertial range; even when the scale is doubled,
$a_2$ is reduced only by a factor 0.88.
Hence the variation of $a_2$ may be too small to be observed
in the limited interval of inertial range.
In this situation we conclude that $a_2$ decreases
as $r$ increases from the dissipative to inertial scale.

\begin{figure}
\psfig{file=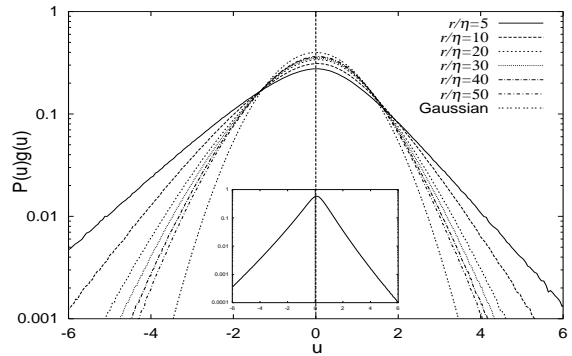,width=75mm,height=48mm}
\caption{$P(u)g(u)$ vs $u$ for various values of $r$.
The dotted line is the normalized Gaussian with $g(u)=1$.
The inset is $P(u)$ for $r/\eta=5$, where the stretched exponential
can be sensed for intermediate amplitudes.}
\label{fig:3}
\end{figure}

\noindent
(iii) $P(u)g(u)$.  The expression (\ref{P}) suggests that the combined form
$P(u)g(u)$ is simpler than $P(u)$ itself.
Figure \ref{fig:3} is a collection of
$P(u)g(u)$ with varying $r$, where the normalization $\int g(u)P(u) du=1$ is
employed.   The Gaussian part at small amplitudes and the exponential-like
part at larger amplitudes are perceptible,
although the slope of the tail becomes
steeper as $r$  increases.
The PDF is obtained by dividing $P(u)g(u)$ by $g(u)$.
The inset of Fig.~\ref{fig:3} is the PDF for $r/\eta=5$, which exhibits
the stretched exponential $\exp[-x^{\rho}]$ with $\rho<1$.
Hence the difference between the exponential distribution in $P(u)g(u)$ and
the stretched exponential one in $P(u)$ is
due to the prefactor $1/g(u)$;
it is checked directly by using the fitting formula for $r/\eta=5$,
although the result is not shown here.
It should be emphasized that the prefactor is important only for small
and intermediate amplitudes, and its effect tends to vanish as $u$
becomes large.

\begin{figure}[t]
\psfig{file=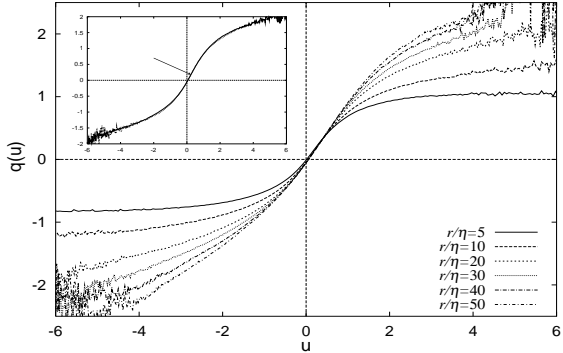,width=75mm,height=48mm}
\caption{$q$ vs $u$ for various values of $r$.
In order to see the antisymmetry of $q(u)$ about a point
marked by an arrow,
we depicted the curve for $r/\eta=20$ (a solid line) and the inverted curve
(a dotted line)
about the point in the inset.  The difference is not appreciated. }
\label{fig:4}
\end{figure}

\begin{minipage}{0.45\textwidth}
\begin{table}
\caption{The fitting parameters $a_0, -a_1$ and $a_2$ for various values of
$r$ in the region $-5 \leq u \leq -0.2$.\label{table:1}}
 \begin{tabular}{ccccccc}
 $r/\eta$ & 5 & 10 & 20 & 30 & 40 & 50 \\
 \tableline
 $a_0$ & 0.40 & 0.52 & 0.67 & 0.75 & 0.85 & 0.88 \\
 $-a_1$ & 0.59 & 0.60 & 0.43 & 0.33 & 0.18 & 0.14 \\
 $a_2$ & 0.26 & 0.10 & 0.048 & 0.039 &  0.054 & 0.050
 \end{tabular}
 \end{table}%
\end{minipage}
\noindent
(4) $q(u)=h(u)/g(u)$.  The ratio function $q(u)$ determines the skeleton of
the PDF.  Since
\begin{equation}
  \frac{H(U_r)dU_r}{G(U_r)}
      =\frac{U_r\av{T(U_r)|U_r}}{\nu G(U_r)}\frac{dU_r}{U_r},
\end{equation}
it is reasonable to think that $q(u)$ is a ratio of the energy
transfer with fixed $u_r$ to the spatial averaged dissipation rate with
fixed $u_r$.  Figure \ref{fig:4} is the curve $q(u)$ for
various  values of $r/\eta$.
For $r/\eta=5$, it     looks the shifted $\tanh(u)$.
As $r$ increases the curves are gradually deviated from $\tanh(u)$,
which is consistent with the fact
that $g(u)$ is dominated more weakly by the term $a_2 u^2$ since $a_2$
decreases with $r$.  It is imagined, however, that as the system size goes
to infinity, $q(u)$ approaches a constant value for large $u$,
implying the exponential tail for $P(u)$.

The PDF of the longitudinal velocity increment is skewed, and its effect is
reflected in the property that $q(u)$ is not antisymmetric about the origin,
as seen from Fig.~\ref{fig:4}.
It is interesting to notice,
however, that the curves $q(u)$ vs $u$ are antisymmetric
about a point marked by an arrow.
In fact, the inverted curve (a dotted line)
about the marked point in the inset of Fig.~\ref{fig:4}
coincides almost with the original one.
When the point is denoted as $(n, m)$, $n \sim m \sim 0.2$ for any separation.
Under this situation the function $q(u)$ can be written as $q(u)=f(u-n)+m$,
where $f(u)$ is an antisymmetric function.

A conjecture for such an expression is as follows.  Let us start with an
antisymmetric function $f(u)$:
\begin{displaymath}
    P(u)g(u)=C \exp \left[ -\int^{u}_0 f(x)dx \right].
\end{displaymath}
Since the third order moments of $u$ (skewness) must be negative
because of the energy transfer, we must have a skew $P(u)$.
($g(u)$ is approximately symmetric around the
region of $u$ which we are concerned.)
The simplest way to do so is to multiply the above expression by $e^{-m u}$.
However the obtained PDF yields the negative first order moment.
In order to make the first order moment zero, we must shift $u$
by a certain amount $n$ as
\begin{displaymath}
    P(u)g(u)=C {\rm e}^{-mu} \exp \left[ -\int^{u-n}_{-n} f(x)dx \right],
\end{displaymath}
which agrees with the observation. Unfortunately we have not known a reason
for the conjecture.

The present result is summarized for the PDF except for the prefactor effect
signified by $1/g(u_r)$.
For simplicity we ignore its asymmetric effect.
First we consider the limit of the infinite Reynolds number.
In the region with small values of $u_r$ the PDF is Gaussian;
$P(u_r) \sim $ ${\rm e}^{-\alpha(r) u_r^2}$,
where $\alpha(r)=c_0(r)/(2a_0(r))$ is a
decreasing function of $r$.  For large values of $u_r$
$P(u_r) ={\rm e}^{-\beta(r)u_r}$,
where $\beta(r)=c_1(r)/a_2(r)$ is an increasing function of $r$.
The transition value $v_1(r)$ is
of order of $\beta(r)/\alpha(r)$,
which is an increasing function of $r$.  Around $u_r \sim v_1(r)$ the
PDF transits from Gaussian to exponential.

Now we turn to the case of finite Reynolds numbers.
I n this case large amplitudes of $u_r$ with a fixed $r$
cannot exist beyond the cutoff amplitude $v_2(r)$,
which increases with the Reynolds number.
Then we have the following PDF.
For $v_1(r)<u_r<v_2(r)$ it is of exponential type.
When $r$
increases with the Reynolds number fixed, $v_1(r)$ becomes of the same
order as $v_2(r)$ at $r=r_0$.
Then for $r>r_0$  the PDF never becomes
exponential.  If the Reynolds number is
small enough,
the situation $v_2(r)<v_1(r)$ occurs for any $r$, implying that the
exponential distribution cannot be observed,
in agreement with the observation that
the exponential distribution was identified only in high Reynolds
numbers\cite{rf:po,rf:cm}.

When we consider the PDF in a fully developed turbulence, the following
questions arise:
what is the PDF of     smaller scales when the Reynolds number
increases?
Is it the stretched exponential as ${\rm e}^{-x^{\rho}}$
with $\rho$ going to zero?
The present study answers these questions on the basis of the
physical argument supported by the simulation.
The answer is  that the eventual PDF is the exponential for
large amplitudes.
Only the slope increases with $r$.
The stretched exponential
is spurious due to the prefactor and it vanishes for large amplitudes.


\end{document}